\documentclass[aip,jcp,twocolumn,superscriptaddress,10pt]{revtex4-1}
\usepackage{color}
\usepackage{fancyhdr} 
\usepackage{varwidth} 
\usepackage{graphicx,subfigure}  
\usepackage{bm} 
\usepackage{listings}
\usepackage{enumerate}%
\usepackage[colorlinks,citecolor=blue,linkcolor=blue]{hyperref}
\usepackage{amsmath}
\usepackage{threeparttable}
\usepackage{graphicx}
\usepackage{amssymb}
\usepackage{amsthm}
\usepackage{rotating}
\usepackage{multirow}

\makeatletter
\def\@email#1#2{%
 \endgroup
 \patchcmd{\titleblock@produce}
  {\frontmatter@RRAPformat}
  {\frontmatter@RRAPformat{\produce@RRAP{*#1\href{mailto:#2}{#2}}}\frontmatter@RRAPformat}
  {}{}
}%
\makeatother

\begin{document}
\title{Dynamic surface tension of the pure liquid-vapor interface subjected to the cyclic loads}

\author{Zhiyong Yu}
\affiliation{State Key Laboratory of Precision Spectroscopy, School of Physics and Electronic Science, East China Normal University, Shanghai 200241, China}
\author{Songtai Lv}
\affiliation{State Key Laboratory of Precision Spectroscopy, School of Physics and Electronic Science, East China Normal University, Shanghai 200241, China}
\author{Xin Zhang}
\affiliation{State Key Laboratory of Precision Spectroscopy, School of Physics and Electronic Science, East China Normal University, Shanghai 200241, China}
\author{Hongtao Liang}
\thanks{lianght@zjlab.ac.cn}
\affiliation{Research and Development Department, Zhangjiang Laboratory, Shanghai 201204, China}
\author{Wei Xie}
\thanks{weixie4@shu.edu.cn}
\affiliation{Materials Genome Institute, Shanghai University, Shanghai 200444, China}
\author{Yang Yang}
\thanks{yyang@phy.ecnu.edu.cn}
\affiliation{State Key Laboratory of Precision Spectroscopy, School of Physics and Electronic Science, East China Normal University, Shanghai 200241, China}

\begin{abstract}
We demonstrate a methodology for computationally investigating the mechanical response of a pure molten lead surface system to the lateral mechanical cyclic loads and try to answer the question: how dose the dynamically driven liquid surface system follow the classical physics of the elastic-driven oscillation? The steady-state oscillation of the dynamic surface tension under cyclic load, including the excitation of high frequency vibration mode at different driving frequencies and amplitudes, was compared with the classical theory of single-body driven damped oscillator. Under the highest studied frequency (50 GHz) and amplitude (5\%) of the load, the increase of the (mean value) dynamic surface tension could reach $\sim$ 5\%. The peak and trough values of the instantaneous dynamic surface tension could reach (up to) 40\% increase and (up to) 20\% decrease compared to the equilibrium surface tension, respectively. The extracted generalized natural frequencies and the generalized damping constants seem to be intimately related to the intrinsic timescales of the atomic temporal-spatial correlation functions of the liquids both in the bulk region and in the outermost surface layers. These insights uncovered could be helpful for quantitative manipulation of the liquid surface tension using ultrafast shockwaves or laser pulses.

\end{abstract}
\maketitle

\section{Introduction}
Variation of the surface tension of molten metal plays decisive role in additive laser manufacturing and powder metallurgy\cite{Khairallah16,Korobeinikov21}. Knowledge of the chemical physics of the dynamic evolution of surface tension is critical in tuning the capillary phenomenon\cite{Sheng14,Girot19,Rossello21} and critical to many advanced processing and manufacturing technologies. Due to the difficulties in direct experimental measurement of the fast evolving dynamic surface tension of liquids\cite{Hauner17}, atomistic simulations plays an important role in understanding the microscopic mechanism of the dynamic behaviors of liquids, yet the number of such simulation studies are quite limited\cite{Lukyanov13,Baidakov19}. 

The current study is motivated by recent demonstrations of ultrafast manipulation of shape and kinetics of condensed matter interfaces via implantation of energy packets\cite{Zalden16,Yang20,Wei21,Wu22}, and especially by the atomistic simulation study by Li et al.\cite{Li22} on ultrafast modulation of the molten metal dynamic surface tension (variation magnitudes could reach over one-fifth of their equilibrium values) within picoseconds under femtosecond laser single-pulse irradiation. Li et al. found that the laser irradiation-induced shockwave results in a significant and biased adjustment in atomic packing density and finally leads to the ultrafast variation in the surface stress distribution along the dynamic molten metal surfaces. Such significant and ultrafast atomic femtosecond laser-induced density oscillation has been detected experimentally with ultrafast electron diffraction technique \cite{Wu22} earlier in the same year. However, more insights into the ultrafast dynamics of liquid surface in response to the various extreme conditions and applied loads\cite{Zellner07,Tsuji07,Chen12} are urgently needed.

Liquid surface in or near equilibrium is widely deemed as analogous to an elastic membrane film while interpreting capillary wave fluctuations\cite{Rowlinson82} and curvature-dependent surface tension variations\cite{Safran03,Ma21}. However, far from equilibrium systematic evaluations of the mechanical response of dynamic liquid surface tension under extreme conditions is rare. To what extent a dynamically driven liquid surface system under extreme loads on the timescales of picoseconds or shorter follows classical elastic physics remains an open question.

In this study, we have carried out atomistic simulations of the dynamics of pure Pb liquid surfaces subjected to lateral mechanical cyclic loads to investigate the aforementioned question. The dynamic surface tension in response to the load was found to follow mostly the theory of the driven damped oscillator in classical mechanics textbooks, yet two clear distinctions were observed, which were caused by complex adjustment of the particle-packing near the liquid surface. In addition, the two generalized elastic properties extracted in our study, i.e., the natural frequencies and the damping constants, are discussed to be linked with the intrinsic timescales regarding atomic temporal-spatial correlation functions of liquids.

\section{Simulation Methods}

We focus on the molten Pb surface (or liquid Pb-vapor interface, LVI) at melting point temperature, $T_\mathrm{m}$. The current MD simulations employ Landa et al.'s embedded-atom-method (EAM) potential for Al-Pb alloy.\cite{Landa00}. The melting point of Pb, $T_\mathrm{m}=615.2$ K, predicted in the crystal-melt coexistence simulation\cite{Yang12}, was consistent with the experimental values of 600.7 K. This EAM potential has been employed in the exploration of the microscopic structure and thermodynamics properties of the Pb liquid phase interfacial systems, such as the (solid)Al-(liquid)Pb interfaces or the (liquid)Al-(liquid)Pb interfaces\cite{Yang12,Yang13,Yang14,Liang18}, yielding predictions of both the solid-liquid interfacial roughening transition temperature\cite{Yang13} and the excess line free energy of the steps at the faceted (solid)Al-(liquid)Pb interface\cite{Liang18}, in excellent agreement with the in-situ transmission electron microscopy measurements\cite{Gabrisch01}.

\begin{figure}[!htb]
\centering
\includegraphics [width=0.48\textwidth] {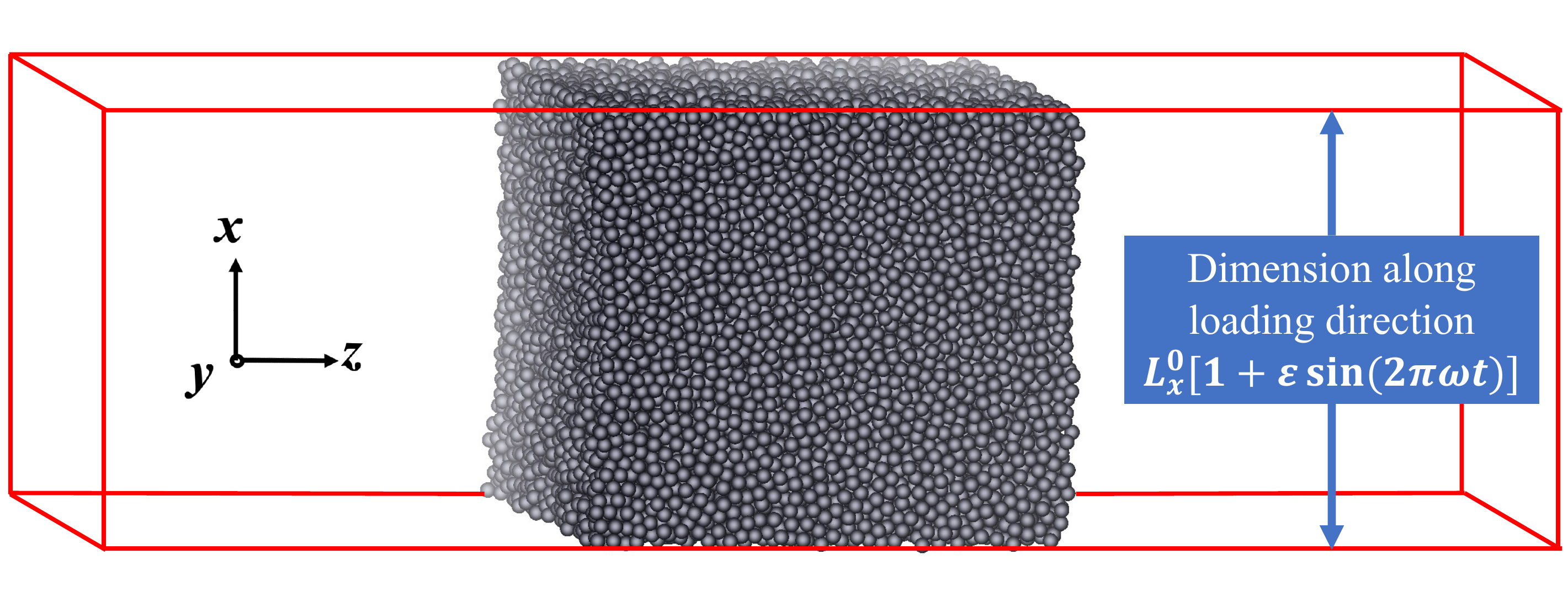}
\caption{Illustration of the (non-)equilibrium MD simulation setup of the Pb LVIs at melting point temperature. The simulation box contains the melt and vapor phases, and two LVIs parallel to the $xy$ plane. The non-equilibrium MD simulations are initiated by applying cyclic loads (Eq.\ref{eq1}) along $x$ axis.}
\label{fig1}
\end{figure}

The MD simulations in this study are performed utilizing LAMMPS\cite{Plimpton95}, with the time-step set as 1fs. All simulations for studying the dynamic surface tensions follow an equilibrium liquid-vapor surface system performed in the canonical ensemble (with constant $NVT$). Periodic boundary conditions (PBC) are used in $x$, $y$, and $z$ dimensions. The dimensions of the simulation box are 100 \AA $\times$ 100 \AA $\times$400 \AA. A liquid slab of around 100 \AA \ in thickness, containing 32,800 liquid Pb atoms, is placed at the center position of the simulation box along $z$ axis, generating two LVIs across the simulation box, as seen in Fig.\ref{fig1}. The equilibrium temperature is set at $T=T_\mathrm{m}=615.2$ K using Nos\'e-Hoover thermostat\cite{Yang12}. The $NVT$ simulations are performed for over 50 ns to ensure the LVIs are fully relaxed into their thermodynamics equilibrium state.

Following the equilibrated molten Pb surface system, we initiate the non-equilibrium MD simulations by applying cyclic loads parallel to the surface system to investigate the steady oscillation state dynamic surface tension and calculate the microscopic quantities of the LVIs experiencing cyclic load. In these non-equilibrium simulations, as illustrated in Fig.\ref{fig1}, we apply the cyclic load $f(t)$ along one direction parallel to the LVI. Specifically, the dimension of the simulation box along the $x$-axis is adjusted as $L_x(t)=L_x^0\times f(t)$, in which $L_x^0=100$\AA \ corresponds to the box dimension along $x$ in the equilibrium $NVT$ simulation. The applied cyclic load in the current study follows a simple sinusoidal function,
\begin{equation}
f(t)=1+{\varepsilon}\sin({2\pi\omega}t)=1+{\varepsilon}\sin({2\pi}t/{C}),
\label{eq1}
\end{equation}
where $\varepsilon$ is the cyclic loading amplitude, $\omega=\frac{1}{C}$ is the frequency of the cyclic load, and $C$ is the period time of one load cycle. $f(t)$ is discretely adjusted every 4000 MD steps. That is, in every 4000 MD steps, the atom coordinates are remapped along with the $L_x(t)$. If the box in the $x$ dimension is expanded or contracted, atom coordinates along the $x$ axis would be dilated or concentrated, respectively, to conform to the new box size. In this work, different loading conditions are applied, including four different loading frequencies (50GHz, 25GHz, 5GHz, 1.25GHz) or equivalently, four cyclic periods (20ps, 40ps, 200ps, 800ps), and three different loading amplitudes(1\%, 3\%, 5\%), see in Table.\ref{tab1}.

Note that, realizations of applying cyclic strain or stress loads to the metallic systems with the aid of state-of-the-art atomistic simulations, have been achieved in several previous studies, yielding useful insights, for examples, on interfacial kinetics\cite{Mishin10}, and on the atomic nature of the solid-state fast mechanical relaxations\cite{Zella22}.

\quad
 \par
\quad

Considering the significant collective/streaming velocities of the liquid atoms subjected to cyclic loads, we thermostat the non-equilibrium system experiencing the cyclic load with a layered thermostat technique, which has been employed to address the realistic high thermal conductivity in metals\cite{Yang18}, thus eliminating the potential artificial local heating/cooling from the homogenous thermostat technique\cite{Monk09}. Throughout each non-equilibrium MD simulation, we freeze the center of mass of the liquid slab. The simulation box is sub-divided into bins of thickness 8\AA, parallel to the LVIs (or the $xy$ plane), and the particles within each slab are independently thermostatted at $T=$615.2K.

\begin{table}[!htp]
\caption{Summary of the non-equilibrium MD simulations of the molten Pb surfaces subjected to the cyclic loads Eq.(\ref{eq1}), including the cyclic loading amplitude $\varepsilon$, frequency $\omega$ of the load, and the period time of one load cycle $C$. Also listed include the total time of the cyclic loading simulation $t_\mathrm{NEMD}$, the simulation time after the system reached the steady oscillation state regime $t_\mathrm{steady}$, the number of cycles $n_\mathrm{cyc}$ selected during $t_\mathrm{steady}$ for calculating dynamic surface tensions, and the time length of the transient regime $t_\mathrm{trans}$.}
\begin{center}
\begin{threeparttable}
\begin{tabular}{ccccccccc}
\hline
\hline
${\varepsilon}$&$C$ &$\omega$&$t_\mathrm{NEMD}$ & $t_\mathrm{steady}$ & $n_\mathrm{cyc}$ & $t_\mathrm{trans}$\\
$\%$ & ps & GHz & ns & ns & - & ns\\
\hline
1&20&50		&392&299&1200&93\\
3&20&50		&393&304&1200&89\\
5&20&50		&415&322&1200&93\\
1&40&25		&160&110&550  &50\\
3&40&25		&178&112&600  &66\\
5&40&25		&154&111&600  &43\\
1&200&5		&230&196&160  &34\\
3&200&5		&234&192&150  &42\\
5&200&5		&247&196&120  &51\\
1&800&1.25	&246&156&60    &90\\
3&800&1.25	&354&168&60    &186\\
5&800&1.25	&294&158&60    &136\\
 \hline
 \hline
 \end{tabular}
\label{tab1}
\end{threeparttable}
\end{center}
\end{table}

\section{Calculation Methods}

This section details the methods used for calculating the key thermodynamic quantities across the liquid-vapor interface experiencing the cyclic load after already entering the steady oscillation state, including the calculation methods for obtaining the dynamic interfacial density, stress profiles, and dynamic surface tension, using data from non-equilibrium MD simulations.

\subsection{Dynamic interfacial profiles}

The dynamic interfacial profiles as the functions of delay times over the load cycles, e.g., density profiles, pressure components profiles, and stress profiles\cite{Li22} are firstly calculated.

The dynamic fine-grained density profile across the molten Pb surface,  $\rho(z,\tilde{t^1})$, at the delay time over one load cycle ($\tilde{t^1}\equiv t \mod C$, the superscript ``1'' stands for one load cycle), is computed as the average number of atoms in each discrete bin of spacing $\delta z$ (chosen as $\delta z$=0.1\AA) divided by the volume of the bin, $A\delta z$, where $A$ is the cross-section area,
\begin{equation}
\rho(z,\tilde{t^1})=\frac{\left\langle N_{z}(\tilde{t^1})\right\rangle_{n_\mathrm{cyc}}}{A\delta z},
\label{eq3}
\end{equation}
where $N_{z}(\tilde{t^1})$ is number of particles in the discrete bin at $\tilde{t^1}$, $\left\langle ...\right\rangle_{n_\mathrm{cyc}}$ averages over samples from $n_\mathrm{cyc}$ load cycles of the steady oscillation state, see in Table.\ref{tab1}.

The determination of the stress (and pressure) tensor uses the virial method and subtracts the component due to any local collective/streaming velocities\cite{Todd95}. This definition is applied to map out the dynamic stress fields of the non-equilibrium liquid surfaces subject to rapid expansion (or contraction) along $x$ and rapid contraction (or expansion) along $z$. The dynamic fine-grained pressure components profiles along the surface normal, $p_{\alpha \beta} (z,\tilde{t^1})$, are determined in fine-graining $z$ axis with bin size $\delta z$, and calculated as the sum of the negative per-particle stress tensors $s_i^{\alpha \beta}$ divided by bin volume and the summation run over $N_{z}(\tilde{t^1})$ particles located between $z$ and $z + \delta z$,
\begin{equation}
p_{\alpha \beta}(z,\tilde{t^1})=-\frac{\left\langle \sum_{i} ^{N_{z}(\tilde{t^1})} s_i^{{\alpha \beta}} (\tilde{t^1}) \right\rangle_{n_\mathrm{cyc}} }{A \delta z}.
\label{eq4}
\end{equation}
The dynamic fine-grained stress profile $S(z,\tilde{t^1})$ is defined as the difference between the dynamic fine-grained normal $p_\mathrm{zz}(z,\tilde{t^1})$ and transverse components $\frac{1}{2} \left[ p_{xx} (z,\tilde{t^1}) + p_{yy} (z,\tilde{t^1}) \right]$ of the pressure tensor.

\subsection{Dynamic surface tension}

\begin{figure}[!htb]
\centering
\includegraphics [width=0.48\textwidth] {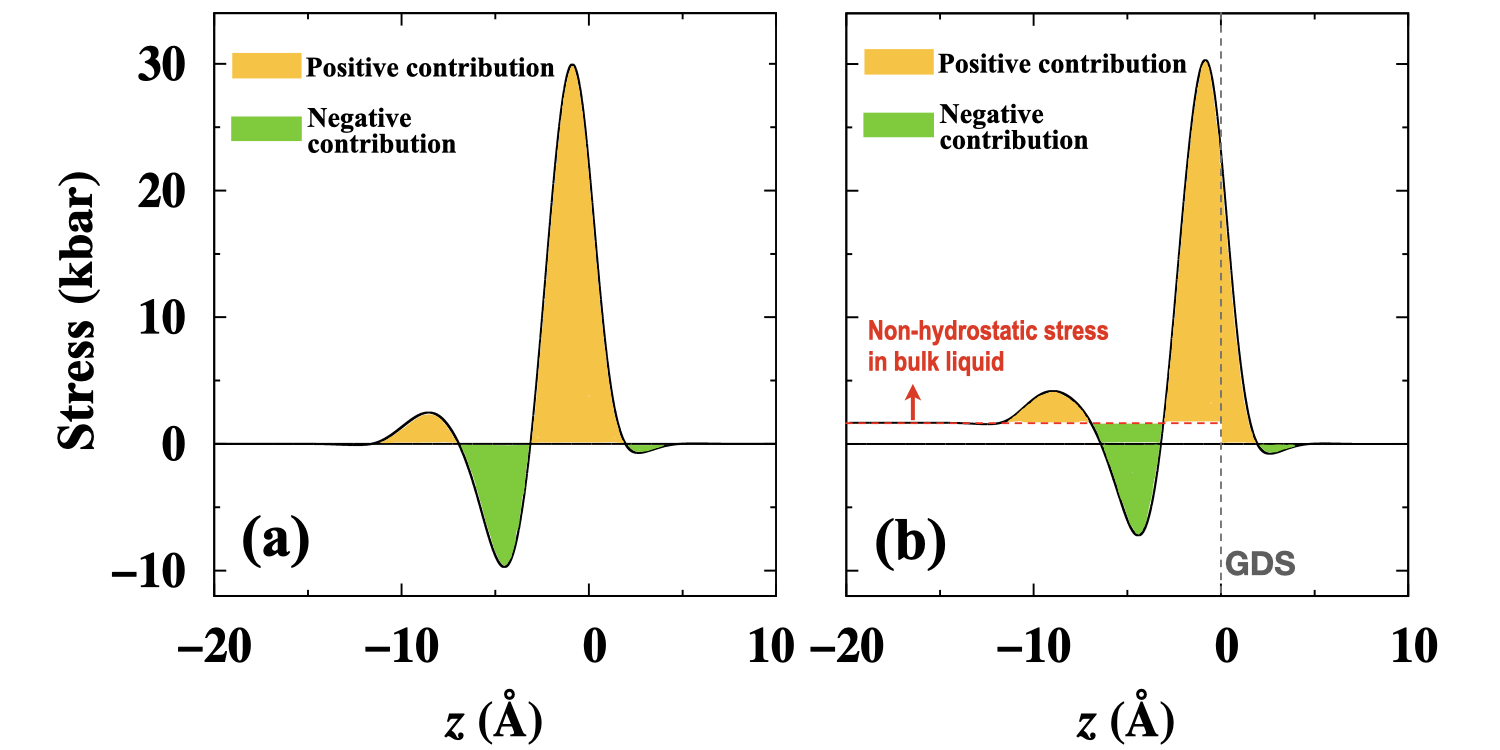}
\caption{Schematic diagrams for the equilibrium stress profile $S_\mathrm{eq}(z)$ of an equilibrium LVI under hydrostatic state (a) and the dynamic stress profile $S(z,\tilde{t^1})$ of a non-equilibrium LVI experiencing the cyclic load and thus under non-hydrostatic tension state (b). In (a), the surface tension calculation is independent of the Gibbs dividing surface (GDS) because the hydrostatic condition applies away from the surface. In (b), the surface tension calculation depends on the position of the GDS (vertical dashed line). The yellow and green shaded areas in both panels denote the positive and negative contributions in implementing Eq.(\ref{eq8a}) and Eq.(\ref{eq8b}), respectively.}
\label{fig2}
\end{figure}

For an equilibrium liquid-vapor interfacial system in which hydrostatic condition applies\cite{Evans74}, the calculation of the surface tension, $\gamma_\mathrm{eq}$, uses the Kirkwood-Buff equation -- the mechanical definition of the surface tension (or excess stress)\cite{Kirkwood49},
\begin{equation}
\gamma_\mathrm{eq} = \int_{z_\mathrm{lo}}^{z_\mathrm{hi}} S_\mathrm{eq}(z) \mathrm{d} z.
\label{eq8a}
\end{equation}
Lukyanov and Likhtman utilized this definition to study dynamic surface tension of a non-equilibrium liquid droplet\cite{Lukyanov13}. 

However, as one can find in the proceeding section, that the applied cyclic loads could modify the hydrostatic condition in the bulk liquid, especially for those cases with larger $\varepsilon$ and $\omega$. To count the dynamic surface tension for highly dynamical states, the vapor phase coexists with the bulk liquid phase in the homogeneous non-hydrostatic condition induced by the applied load. In contrast to the above mentioned liquid surface system under a hydrostatic equilibrium state, the calculation of the surface tension (or excess stress) for this type of interfacial systems depends on the position of the Gibbs dividing surface (GDS), and the mechanical definition of the surface tension Eq.(\ref{eq8a}) is not applicable\cite{Frolov10}, as seen in Fig.\ref{fig2}(b).

The GDS position at $\tilde{t^1}$, for each dynamic liquid Pb surface in the steady oscillation state, is chosen such that the excess number of particles (in each $\rho(z,\tilde{t^1})$) equals to zero, i.e., 
$N_{\text {excess }}(\tilde{t^1}) = N - \rho_l(\tilde{t^1}) AL_l(\tilde{t^1})  - \rho_v(\tilde{t^1})AL_v(\tilde{t^1}) =0$, where $\rho_v(\tilde{t^1})$ and $\rho_l(\tilde{t^1})$ are the number densities in the bulk vapor phase and bulk liquid phase, respectively. $L_v(\tilde{t^1})$ and $L_l(\tilde{t^1})$ are the corresponding lengths along $z$ of the bulk vapor phase and bulk liquid phase, defined by the GDS at $\tilde{t^1}$, respectively.

With the knowledge of $L_v(\tilde{t^1})$ and $L_l(\tilde{t^1})$, the dynamic surface tension (or the interfacial excess stress) is thus calculated as,
\begin{equation}
\gamma(\tilde{t^1})=\left[\int_{z_\mathrm{lo}}^{z_\mathrm{hi}} S(z,\tilde{t^1}) \mathrm{d} z\right]-S_l(\tilde{t^1})L_l(\tilde{t^1}),
\label{eq8b}
\end{equation}
in which, $S_l(\tilde{t^1})$ is the finite value of the stress in the homogeneous non-hydrostatic liquid phase experiencing cyclic load at $\tilde{t^1}$, which is measured from averaging approximately one third of the plateau regions in the dynamic fine-grained $S(z,\tilde{t^1})$ profile. Note that $S_v(\tilde{t^1})=0$ in the vapor phase, so that the corresponding term $S_v(\tilde{t^1})L_v(\tilde{t^1})$ is removed from Eq.(\ref{eq8b}). We implement the calculation in Eq.(\ref{eq8b}) by employing the Simpson rule in the numerical integration.

\begin{figure}[!htb]
\centering
\includegraphics [width=0.3\textwidth] {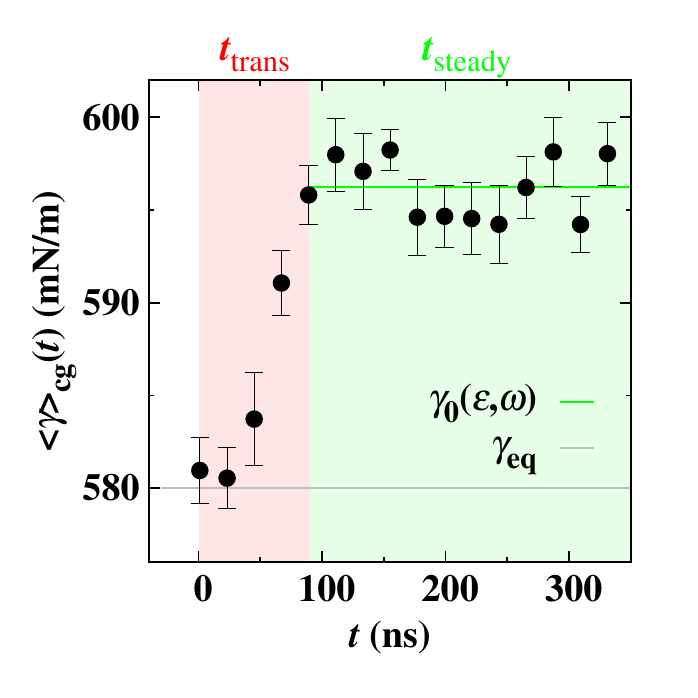}\par
\caption{The resulting coarse-grained dynamic surface tension $\langle\gamma\rangle_\mathrm{cg}$ is plotted as the function of simulation time $t$ for the Pb LVI subjected to cyclic loads with parameters, $\omega=50$GHz, $\varepsilon=3\%$. Here the coarse-grained dynamic surface tension $\langle\gamma\rangle_\mathrm{cg}$ is calculated as the average value of 120 cycles (a total time of 2.4 ns) over every 22 ns, for the LVI experiencing the cyclic load. After an initial transient regime, at around 90 ns, denoted by a vertical line, after which, $\langle\gamma\rangle_\mathrm{cg}$ reaches a constant value suggesting that the response to the applied load enters the steady oscillation state regime. The error bars represent the 95\% confidence intervals estimated from statistical average.}
\label{fig3-}
\end{figure}

\begin{figure*}[!htb]
\centering
\includegraphics [width=0.85\textwidth] {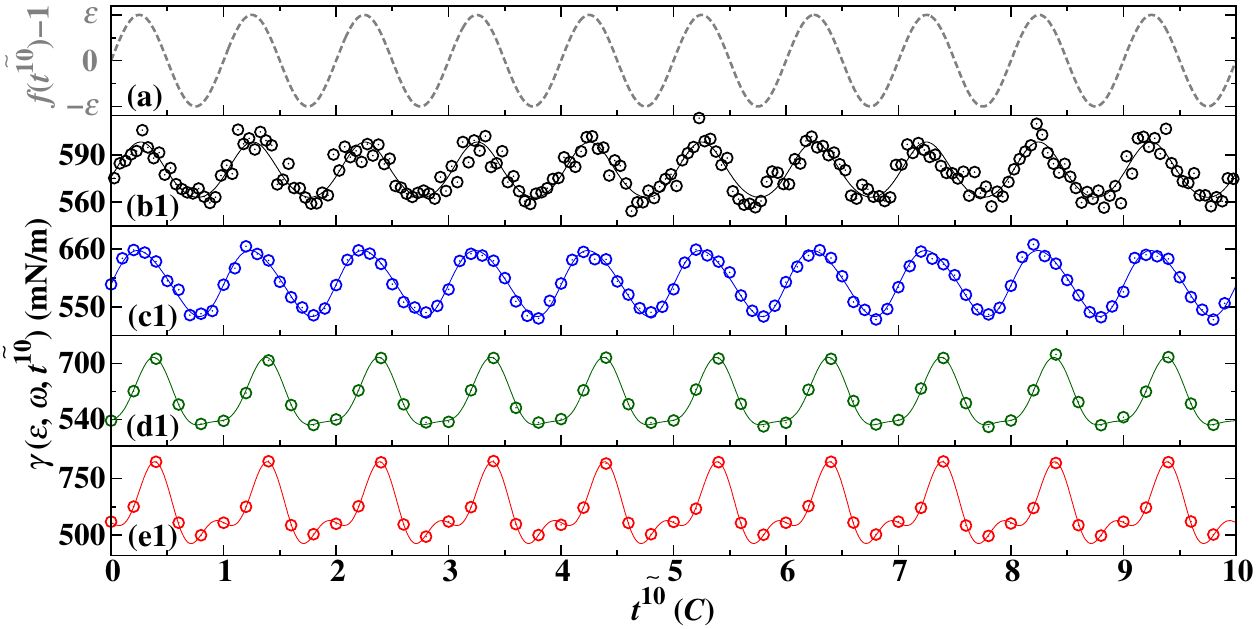}\par
\includegraphics [width=0.85\textwidth] {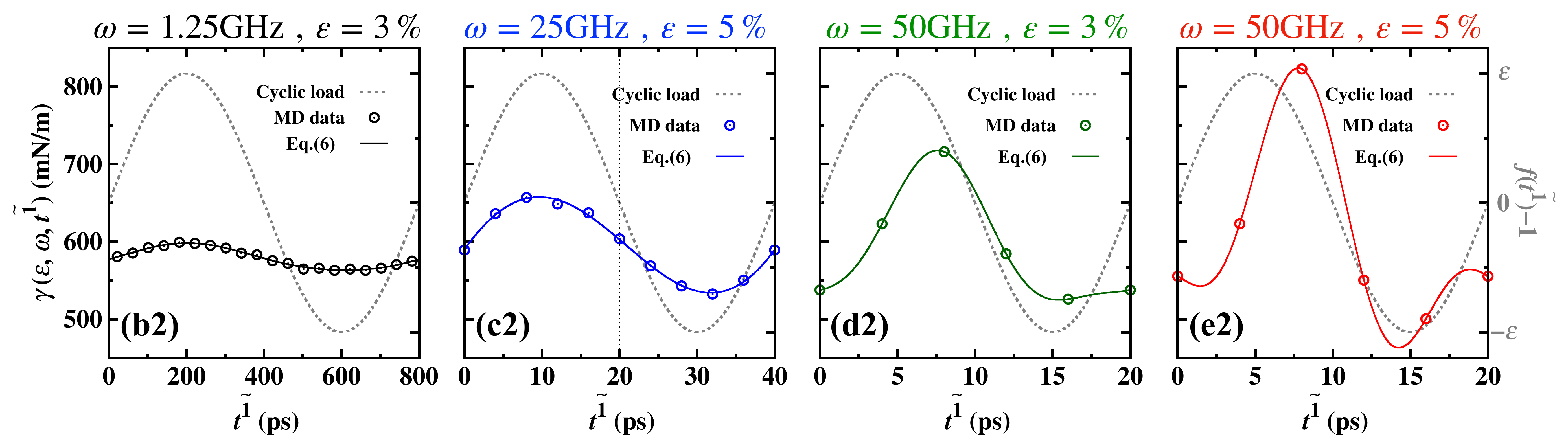}
\caption{Steady oscillation state responses of the dynamic surface tension of the molten Pb surfaces ($T=T_\mathrm{m}=615.2$K) to sinusoidal cyclic loads (gray dashed line in panel (a) and (b2)-(e2)) with different loading frequencies and amplitudes. The statistically averaged dynamic surface tension oscillations are represented with open circles for the initial conditions $\omega=1.25$GHz, $\varepsilon=3$\% (b1,b2); $\omega=25$GHz, $\varepsilon=5$\% (c1,c2); $\omega=50$GHz, $\varepsilon=3$\% (d1,d2); $\omega=50$GHz, $\varepsilon=5$\% (e1,e2). The oscillatory solid lines in panel (b1)-(e1) correspond to the fitting functions using the form of $\frac{\gamma(\varepsilon,\omega,\tilde{t^{10}})}{\gamma_{0}(\varepsilon,\omega)}=1+\sum_{n=1}^{2}A_{n}(\varepsilon,\omega)\sin[2\pi n\omega\tilde{t^{10}}+{\delta}_{n}(\varepsilon,\omega)]$. Similarly, the solid curves in panel (b2)-(e2) correspond to the fitting functions using the form of Eq.(\ref{eq10}). The time axes in these panels are converted to the delay-time $\tilde{t^{10}}$ ($\tilde{t^{10}}\equiv t \mod 10C$) or $\tilde{t^1}$, respectively.}
\label{fig3}
\end{figure*}

\section{results and discussion}
\label{sec-result}

In Fig.\ref{fig3-}, we demonstrate a resulting time evolution of the dynamic surface tension $\langle\gamma\rangle_\mathrm{cg}(t)$ for one LVI system subjected to the cyclic load ($\omega=50$GHz, $\varepsilon=3\%$). The data points plotted in Fig.\ref{fig3-} ($\langle\gamma\rangle_\mathrm{cg}$) correspond to the coarse-grained mean values of 120 cycles (a total time of 2.4 ns) over every 22 ns. After the onset of the cyclic load ($t=0$), the dynamic surface tension of the molten Pb surface increases over a transient regime of around 90 ns. The transient regime is followed by a steady oscillation state regime where the magnitude of $\langle\gamma\rangle_\mathrm{cg}(t)$ converges to a constant value more significant than the equilibrium surface tension $\gamma_\mathrm{eq}=580(2)$ mN/m at $T=T_\mathrm{m}$. The time length of the transient regime $t_\mathrm{trans}$ varies significantly among different simulation cases. See Table.\ref{tab1}. 

We try to learn the dynamic surface tension to the cyclic load with an analogy to the knowledge of the driven damped oscillator in classical mechanics, i.e., the dynamic surface tension of the LVI subjected to the cyclic load is analogous to the instantaneous position of the driven oscillator. A clear difference we notice in the transient regime between the classical mechanical driven damped oscillator and the current investigated system is that the mean position of the oscillator in the former system is predicted to stick to the original equilibrium position as the steady oscillation state is approached. In contrast, the $\langle\gamma\rangle_\mathrm{cg}(t)$ value in the mean dynamic surface tension can rise to a significantly greater value in the dynamic LVI system. This difference could arise because the driven oscillator is a single-body system. In contrast, the LVI system consists of numerous atoms in which the atomistic structure and the mechanical scenario could be substantially rearranged during the transient regime according to the applied loads. It would be necessary to conduct an independent research work on these non-equilibrium microscopic rearrangements. However, such a study is beyond the scope of the current study.

Fig.\ref{fig3}(b1)-(e1) select and present three temporal evolutions of the steady oscillation state dynamic surface tension under sinusoidal cyclic loads (Fig.\ref{fig3}(a)) with different loading amplitudes and frequencies. For the panels (a),(b1)-(e1) of Fig.\ref{fig3}, the scale of the $x$-axis is converted to 10 load cycles, i.e., the dynamic surface tensions $\gamma(\tilde{t^{10}})$ are determined from the statistical averaged dynamic fine-grained stress profile as the functions of $z$ and delay over ten load cycle ($\tilde{t^{10}}\equiv t \mod 10C$, the superscript ``10'' stands for ten load cycles). In these steady oscillation states, the dynamic surface tension oscillates upon a constant baseline value $\gamma_{0}(\varepsilon,\omega)$ with periodic manners following the applied cyclic loads. It is found that the loading amplitude and frequency affect the mechanical responses of the LVIs to cyclic loads. The dynamic surface tension oscillates sinusoidally for those cases with smaller cyclic loading $\omega$ and $\varepsilon$ at precisely the drive frequency. For those cases with more significant cyclic load $\omega$ and $\varepsilon$, additional oscillating components with a period different from the original drive period are found, resulting in an evident deviation from the perfect sinusoidal oscillation (as observed in the cases with more minor $\omega$ and $\varepsilon$). Moreover, as is seen in the panel (b2)-(e2) of Fig.\ref{fig3}, the dynamic surface tension oscillation for the LVI system subjected to the cyclic load changes in its magnitude, phase shift, and the baseline value $\gamma_{0}(\varepsilon,\omega)$ as well. These clues indicate that the response physics of the current LVI system under cyclic load is probably akin to the driven oscillator model theory in classical mechanics. Meanwhile, the dynamic surface tension can be tuned variously through collective modulation of density in the vicinity of the LVI.

Given the obtained (steady oscillation state) results of the dynamic surface tension result, we refer to the Fourier series solution (with two leading terms) for the driven oscillator\cite{Taylor05} in classical mechanics to quantitatively interpret the modulations of the dynamic surface tensions through varying the loading frequency and amplitude. The analytical equation Eq.(\ref{eq10}) is employed to fit the dynamic surface tension results,
\begin{equation}
\frac{\gamma(\varepsilon,\omega,\tilde{t^1})}{\gamma_{0}(\varepsilon,\omega)}=1+
\sum_{n=1}^{2}A_{n}(\varepsilon,\omega)\sin[2\pi n\omega\tilde{t^1}+{\delta}_{n}(\varepsilon,\omega)].
\label{eq10}
\end{equation}
$\gamma_{0}(\varepsilon,\omega)$ in Eq.(\ref{eq10}) is the steady oscillation state constant baseline value, $A_{n}(\varepsilon,\omega)$ and ${\delta}_{n}(\varepsilon,\omega)$ are the resulting oscillation amplitudes and phase difference of the dynamic surface tensions in response to the cyclic load. The integer $n$ denotes the leading two non-constant components in the response function. By fitting the steady oscillation state dynamic surface tension data to Eq.(\ref{eq10}), see in Fig.\ref{fig3} for instance, it is confirmed that our choice of using the theoretical model of the driven oscillator with two leading Fourier series can well satisfied in describing the oscillation in the dynamic surface tension in response of the cyclic load with loading frequency as high as 50GHz and the loading magnitude up to 5\%.

\begin{figure}[!htb]
\centering
\includegraphics [width=0.3\textwidth]{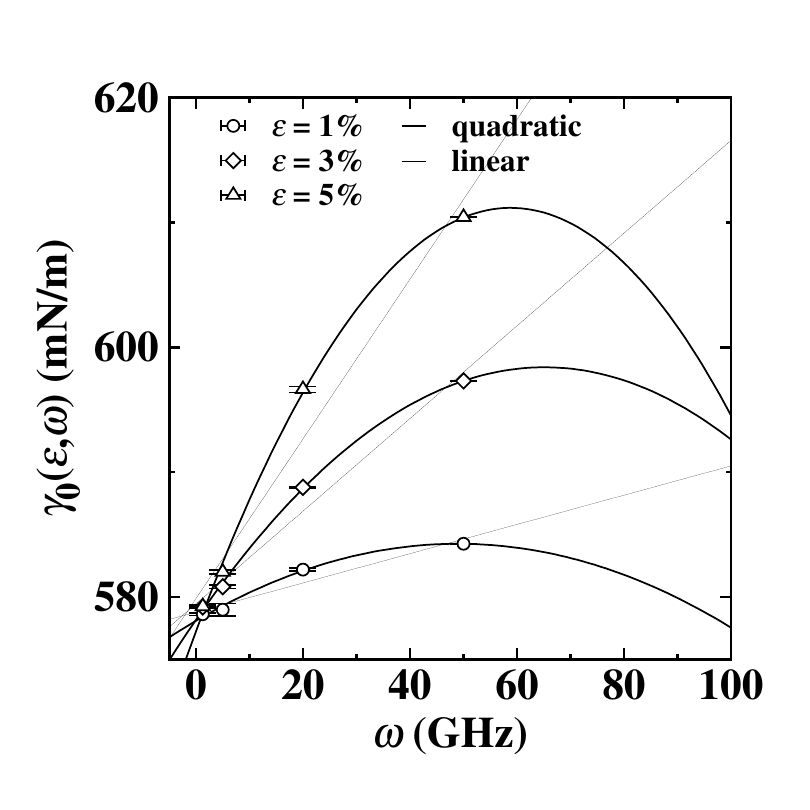}
\caption{Steady oscillation state results of the constant baseline value of the oscillating dynamic surface tension, $\gamma_{0}(\varepsilon,\omega)$, as functions of the cyclic loading frequencies $\omega$ and amplitudes $\varepsilon$. Solid lines plot the weighted least-squares fits to linear and quadratic functions.}
\label{fig5}
\end{figure}

As depicted in Fig.\ref{fig5}, higher loading frequency $\omega$ (or magnitude $\varepsilon$) results in greater values of $\gamma_{0}$ under fixed loading magnitude (or frequency). We fit the data points using linear and quadratic weighted least squares regressions and identify the $\gamma_{0}(\varepsilon,\omega)$ for a fixed $\varepsilon$ follows a quadratically increasing trend in the $\omega$ range less than 50$\sim$60 GHz. As mentioned earlier, the increase of the magnitude of $\gamma_{0}(\varepsilon,\omega)$ could be due to the rearrangement of the atomistic structure and the mechanical scenario in the vicinity of the LVI subjected to the cyclic load. To the best of our knowledge, few theories predict either such atomistic rearrangement or the spatial distribution of the stress along the LVI normal direction. See Ref.\onlinecite{Lu22} and references therein. Nonetheless, more insights will be revealed from the calculated dynamic interfacial profiles in the proceeding context.

\begin{align}
\label{eq11}
A^2_n(\varepsilon,\omega)&=\frac{f^2_n}{{(4\pi^2\omega^2_{0}-4\pi^2n^2\omega^2)^2+16\pi^2\beta^2n^2\omega^2}},\\
\label{eq12}
\delta_n(\omega)&=\arctan\left(\frac{n\beta\omega}{\pi\omega^2_{0}-\pi n^2\omega^2} \right).
\end{align}

In the classical mechanical theory of the driven damped oscillator, the amplitudes $A_n$ and the phase shifts $\delta_n$ (phase differences in the oscillator's motion lags behind the cyclic driving force) for driven oscillations, as the functions of the driving frequency $\omega$, are predicted with the analytical expressions, i.e., Eq.(\ref{eq11}-\ref{eq12}). $f_n$ are the amplitudes of the two leading Fourier components of the cyclic driving force. $\omega_0$ and $\beta$ are the system(material)-dependent natural frequency and the damping constant, respectively. Given the complexity of the current liquid surface system, as compared to the simple single-body oscillator, in the following analysis, we employ different values of the natural frequencies (i.e., $\omega_{01}$, $\omega_{02}$) and the damping constants (i.e., $\beta_1$, $\beta_2$) to interpret the responses of the dynamic surface tension to the cyclic loads, as appears in Eq.(\ref{eq13}-\ref{eq14}).

\begin{align}
\label{eq13}
A^2_n(\varepsilon,\omega)&=\frac{f^2_n}{{(4\pi^2\omega^2_{0n}-4\pi^2n^2\omega^2)^2+16\pi^2\beta_n^2n^2\omega^2}},\\
\label{eq14}
\delta_n(\omega)&=\arctan\left(\frac{n\beta_n\omega}{\pi\omega^2_{0n}-\pi n^2\omega^2} \right).
\end{align}

\begin{figure}[!htb]
\centering
\includegraphics [width=0.48\textwidth] {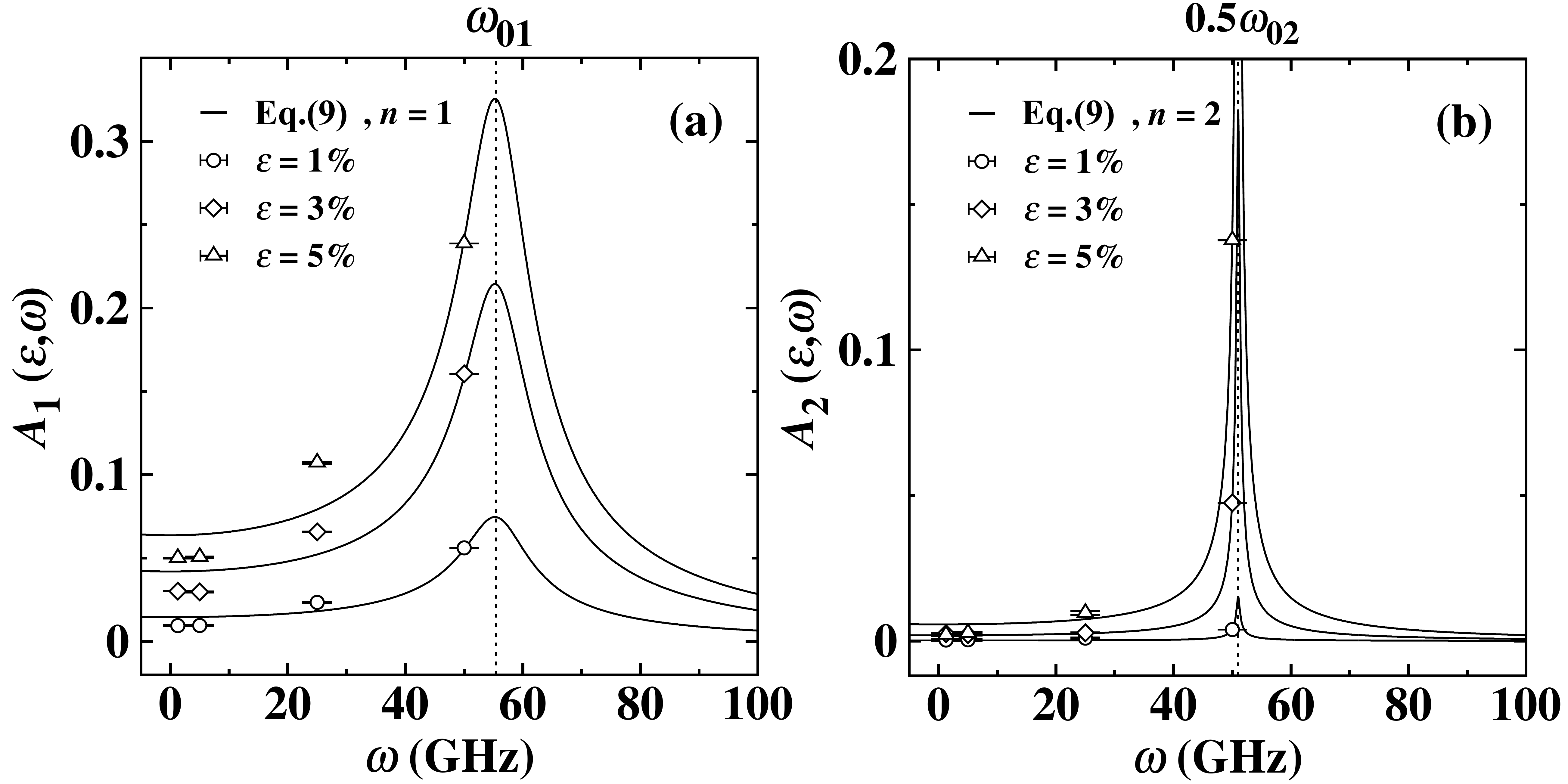}
\caption{The resulting oscillation amplitudes ($A_1$ and $A_2$) of the dynamic surface tensions in response to the cyclic load, as a function of the cyclic loading frequency $\omega$, for three different values of the cyclic loading amplitudes $\varepsilon$. Solid lines in (a) and (b) are the global weighted least-squares fit of all sets of data points to Eq.(\ref{eq13}) with $n=1$ and $n=2$, respectively. The position of the fitted results of natural frequencies (i.e., $\omega_{01}$, $\omega_{02}$) are labeled with the vertical dashed lines.}
\label{fig6}
\end{figure} \par

The two calculated amplitudes ($A_1$ and $A_2$) for driven oscillation of the dynamic surface tensions as functions of the driving frequency ($\omega$) for three different values of cyclic loading amplitude $\varepsilon$ are depicted in Fig.\ref{fig6}(a) and Fig.\ref{fig6}(b). Higher loading frequency and cyclic loading amplitude result in a more significant magnitude of both $A_1$ and $A_2$. Overall, the values of $A_2$ are smaller than that of $A_1$ in most cases. Especially for the case with smaller driving frequency or cyclic loading amplitude, yielding near-zero $A_2$ values and, therefore -- near-perfect sinusoidal oscillations of the dynamic surface tension.

\begin{figure}[!htb]\par
\includegraphics [width=0.48\textwidth] {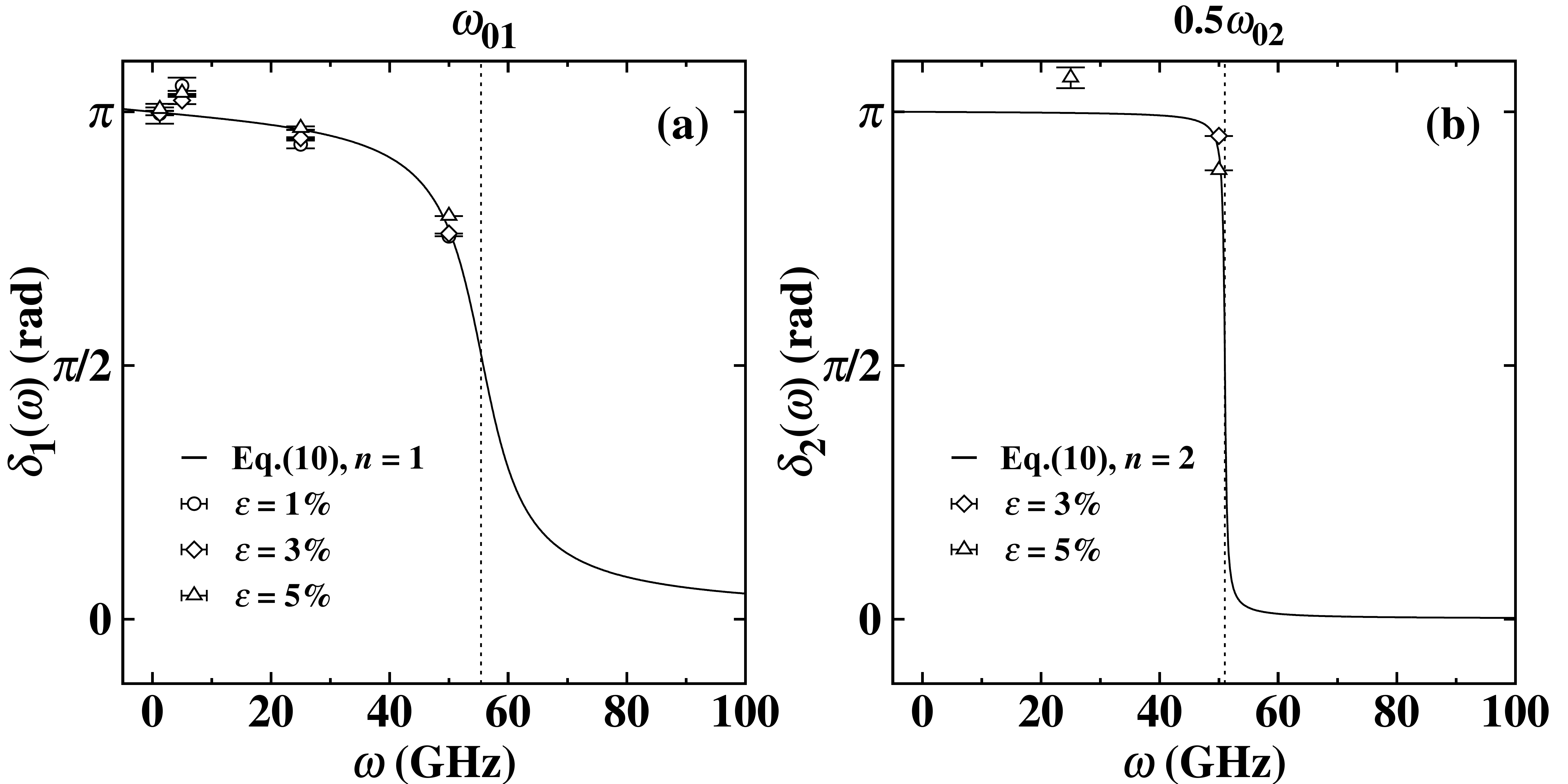}
\caption{The resulting phase shifts relative to the cyclic loads ($\delta_1$ and $\delta_2$) in dynamic surface tensions as functions of the cyclic loading frequency $\omega$ for different $\varepsilon$. Solid lines in (a) and (b) are the global weighted least-squares fit of all sets of data points to Eq.(\ref{eq14}) with $n=1$ and $n=2$, respectively. 
Solid curves in (a) suggest relatively wider resonances than those in (b). Only the cases with finite magnitudes of $A_2$ are reported. The position of the fitted results of natural frequencies (i.e., $\omega_{01}$, $\omega_{02}$) are labeled with the vertical dashed lines.}
\label{fig7}
\end{figure}

The calculated phase shifts relative to the cyclic loads in dynamic surface tensions ($\delta_1$ and $\delta_2$), as functions of $\omega$ and $\varepsilon$ are shown in Fig.\ref{fig7}(a) and Fig.\ref{fig7}(b). Note that, for $\delta_2(\omega)$, only the cases with finite magnitudes of $A_2$ are obtained and reported in Fig.\ref{fig7}(b), whereas the uncertainties of the $\delta_2$ are significantly large for those cases with near-zero $A_2$. It is noticed that the values of $\delta_1$ and $\delta_2$ for different driving frequencies are nearly independent of the cyclic loading amplitude $\varepsilon$, as predicted in Eq.(\ref{eq14}). For the very small $\omega$, $\delta_1$ and $\delta_2$ are close to zero, indicating that oscillations of the dynamic surface tensions are almost perfectly in step with the cyclic load (e.g., the case in Fig.\ref{fig3}(b-c)). As $\omega$ increases, the values of $\delta_1$ and $\delta_2$ increase, yet not reaching a magnitude of $\pi/2$ lag behind the applied cyclic load.

The solid curves in Fig.\ref{fig6} and Fig.\ref{fig7} are weighted least-squares fits to Eq.(\ref{eq13}) and Eq.(\ref{eq14}), respectively. Both equations well fit the entire sets of computed data from NEMD simulations (i.e., $A_1(\varepsilon,\omega)$, $A_2(\varepsilon,\omega)$, $\delta_1(\omega)$ and $\delta_2(\omega)$) with comparable accuracy. The fits of the data give estimates of $\omega_{01}=55.8(8)$GHz, $\omega_{02}=102.0(6)$GHz, $\beta_1=5.5(8)$GHz, and $\beta_2=0.5(1)$GHz. The good quality of the fitting again indicates that the driven oscillation of the dynamic surface tension is nearly consistent with the physics of the driven oscillator in classical mechanics, despite that the current systems require one additional natural frequency and damping constant. 

With these estimations, one could tell the systems are under underdamped conditions, i.e., $\beta_1/\omega_{01}\approx0.1$ and $\beta_2/\omega_{02}\approx0.005$. As the natural frequencies ($\omega_{01}$, $\omega_{02}$) are approached from below, the two amplitudes ($A_1$ and $A_2$) of the driven dynamic surface tension oscillations dramatically increase to their corresponding resonance peaks. Notice the ratio of $\beta_2/\omega_{02}$ is 20 times smaller than the ratio of $\beta_1/\omega_{01}$, suggesting that the second component ($n=2$) of the dynamic surface tension oscillation owns a narrower peak in amplitude and a more abrupt decay in phase shift than the first component ($n=1$), agree with the data shown in Fig.\ref{fig6} and Fig.\ref{fig7}. The highest cyclic loading frequency in the current study, $\omega=50$GHz, is very close to the natural frequency $\omega_{01}$; meanwhile, the $n=2$ component, with frequency $\omega=50$GHz, is nearly equal to half of the natural frequency $\omega_{02}$. These indicate that both component terms are almost at resonance, thus resulting in two strong responses at the same time so that the dynamic surface tensions deviate from the perfect sinusoidal oscillation, e.g., panel (d1,d2) and (e1,e2) in Fig.\ref{fig3}.

 \par

To date, there has been little knowledge on the natural frequencies and damping constants for the surface tension oscillation of the liquid surfaces, which have been usually treated as elastic membrane system\cite{Rowlinson82,Safran03,Ma21}. With
above fitted data ($\omega_{01}$, $\omega_{02}$, $\beta_1$, and $\beta_2$), we carry out the proceeding discussions on the possible nature of these quantities.

Because the liquid surfaces are composed of massive, temporally, and spatially correlated atoms, their dynamic surface tension is ascribed to the changes in the microscopic packing structure, which deviates from the equilibrium packing scenario. We, therefore, firstly speculate that the natural frequencies are related to the relaxation process of the liquid atomic density fluctuation, namely, the spectrum of longitudinal-current fluctuations or the dynamic structure factor. Considering each oscillation period contains two descent and two ascent parts, the fitting results of $\omega_{01}=55.8(8)$ GHz and $\omega_{02}=102.0(6)$ GHz correspond to timescales of around 4.48 ps and 2.45 ps, respectively. The latter timescale (from $\omega_{02}$) is comparable to the characteristic longitudinal collective dynamics timescale 2.12(12) ps. i.e., the bulk liquid density relaxation time, defined as the inverse half-width of the dynamic structure factor\cite{Zhang22}. The former timescale (from $\omega_{01}$) is more significant than twice the bulk liquid density relaxation time. Nevertheless, Reichert et al.\cite{Reichert07} and del Rio et al.\cite{Rio20} reported a drastic slowing down of the longitudinal collective dynamics at near-surface atomic layers, i.e., density relaxation time increased at least by a factor of 2. The above clues well support our speculations on the nature of the natural frequencies of the driven surface tension oscillations and imply that the natural frequencies of the surface tension oscillation may be spatially inhomogeneous, i.e., the near-surface layers and sub-layers may respond differently to the applied load.

As to the damping constants, the two fitted results of $\beta_1=5.5(8)$ GHz and $\beta_2=0.5(1)$ GHz correspond to timescales of around 45 ps and 500 ps, respectively. Again, our preliminary speculation to the nature of these damping constants falls to the timescale of the ``molasses'' decaying tail in the Green-Kubo integrand for the liquid shear viscosity, $\eta=\int_0^{\infty}\eta(t)\mathrm{d}t$\cite{Hansen13}. The $\eta(t)$ is determined from the autocorrelation function of an off-diagonal element of the atomic stress tensor\cite{Berk02}. Unfortunately, our calculation of such decaying time for the bulk molten Pb at $T_\mathrm{m}$ is around 1 ps, which is about two orders of magnitude smaller than the timescale corresponding to $\beta_2$, respectively. Moreover, few studies have uncovered any novelty of the local shear viscosity at liquid surfaces. Therefore, we have not gathered sufficient and firm evidence that could support our preliminary speculation about the nature of the damping constants for the driven oscillation in $\gamma(\varepsilon,\omega,\tilde{t^1})$.

\begin{figure}[!htb]
\includegraphics [width=0.48\textwidth] {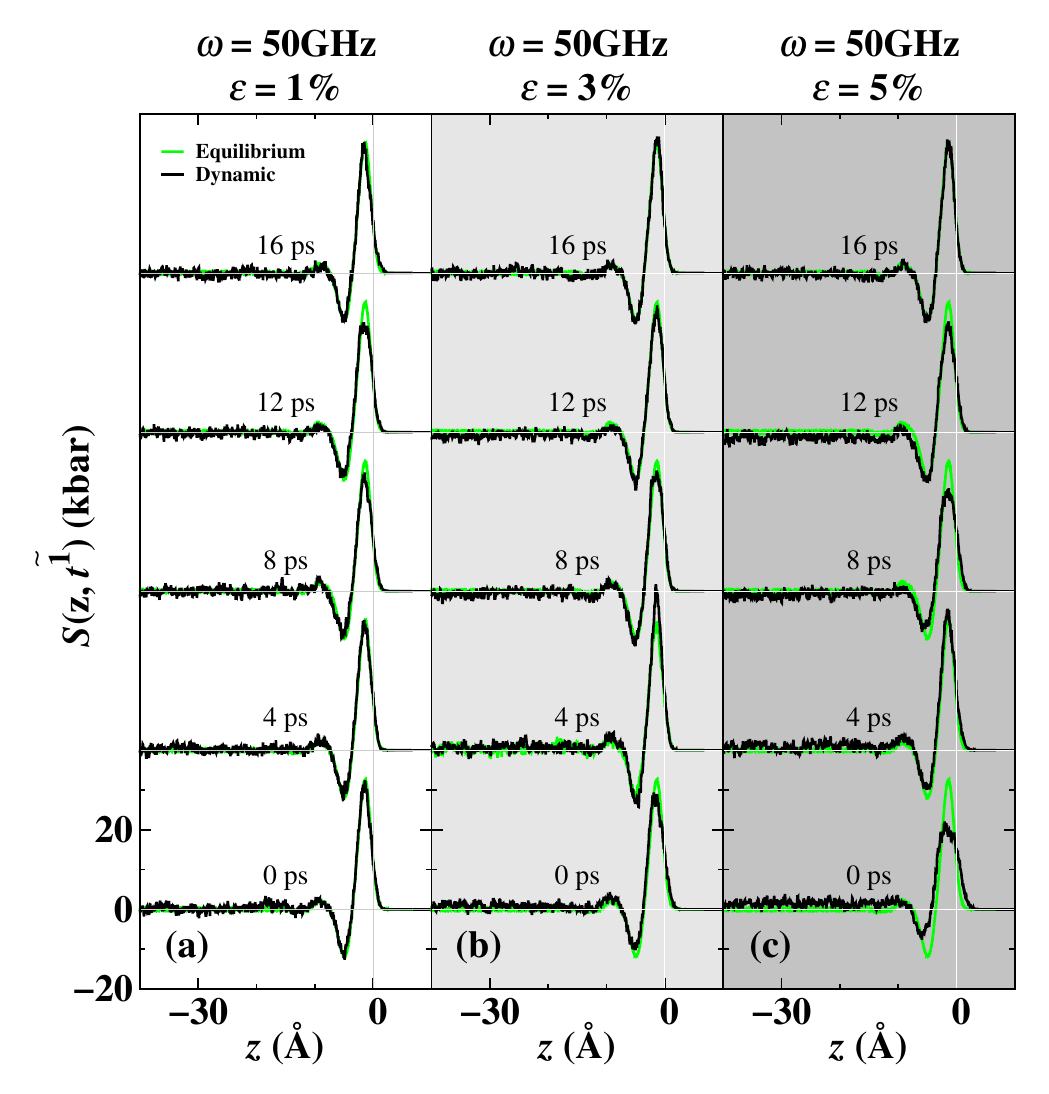}
\caption{The steady oscillation state dynamic fine-grained stress profiles at different delay times over one load cycle ($\tilde{t^1}\equiv t \mod C$). Under the cyclic loading frequency of 50GHz, and loading amplitude of $\varepsilon=1\%$ (a), $\varepsilon=3\%$ (b), $\varepsilon=5\%$ (c). Bottommost and topmost curve represents the result for the $\tilde{t^1}=0 ps$ and $\tilde{t^1}=16 ps$, respectively. The curve in green represents the fine-grained stress profile of the equilibrium molten Pb surface at $T=T_\mathrm{m}$, which serves as a reference in comparison with the dynamic interfacial profiles.}
\label{fig8}
\end{figure}

We next examine the spatial-temporal evolutions of the dynamic fine-grained stress profiles and dynamic fine-grained density profiles across the molten Pb surface subjected to the cyclic load to dig for more microscopic insight to interpret the driven oscillation in the dynamic surface tensions. The dynamic microscopic stress shown in Fig.\ref{fig8} is calculated as the difference between the dynamic transverse pressure component and the dynamic normal pressure component. Zero stress regime indicates that the liquid is under hydrostatic conditions, and positive or negative stresses corresponding to the local liquids are under lateral tension or lateral compression, respectively. A prominent positive stress peak is followed by an oscillatory damping structure (smaller negative and much weakened positive peaks), and stress is zero in the bulk region for the molten Pb surface under an equilibrium state. For the liquid surfaces under lateral cyclic loads, such as the cases with loading frequencies of 50GHz shown in Fig.\ref{fig8}, the temporal evolution of the dynamic fine-grained stress profile shape becomes increasingly volatile as the loading amplitudes increase. Specifically, only slight adjustments in the width and amplitude of the positive surface peak, primarily subjected to the outermost surface layer, as the function of delay-time $\tilde{t^{1}}$ are noticed for the smaller driven amplitude case, e.g., $\varepsilon=1\%$.
In contrast, for the more significant loading amplitude cases, e.g., $\varepsilon=3\%$, and $\varepsilon=5\%$, the stress distribution for the region behind the positive peak exhibit evident adjustments. In addition to the more significant adjustments in the width and amplitude of the outermost positive peak, the development of the finite stress in the bulk liquid is seen together with the weakened sub-surface damping structures. These dynamic adjustments for these higher driven amplitude cases indicate that the sub-surface stresses start to contribute to the variation of the dynamic surface tension. It seems that such contribution might not be identical to the part due to the structural adjustment in the outermost positive stress peak and echoing the $n=2$ component contributing to the dynamic surface tension response function mentioned in earlier and proceeding paragraphs. 

According to the classical density functional theory, the system free energy field is represented with the liquid density and the related functionals\cite{Hansen13b}. The global free energy minima determine the equilibrium atomic packings and the equilibrium density distributions. When the local liquid densities are modified, deviating their equilibrium values due to the applied cyclic load, the system would dynamically re-adjusting the density fields towards a direction where the free energies are lowered. For simulation cases under cyclic loads with very low loading frequencies (e.g., $\omega=$1.25 GHz), even the amplitude of the load reaches 5\%. Because the cyclic load-induced regulation in the local liquid densities is too long to rig the intrinsic particle packing re-adjusting process, the corresponding timescales ($C/4=$200 ps) are around two orders of magnitude longer than the bulk liquid density relaxation time (2.12(12) ps), so that the dynamic density and stress profiles for these cases are nearly identical to the equilibrium surface.

\begin{figure}[!htb]
\includegraphics [width=0.48\textwidth] {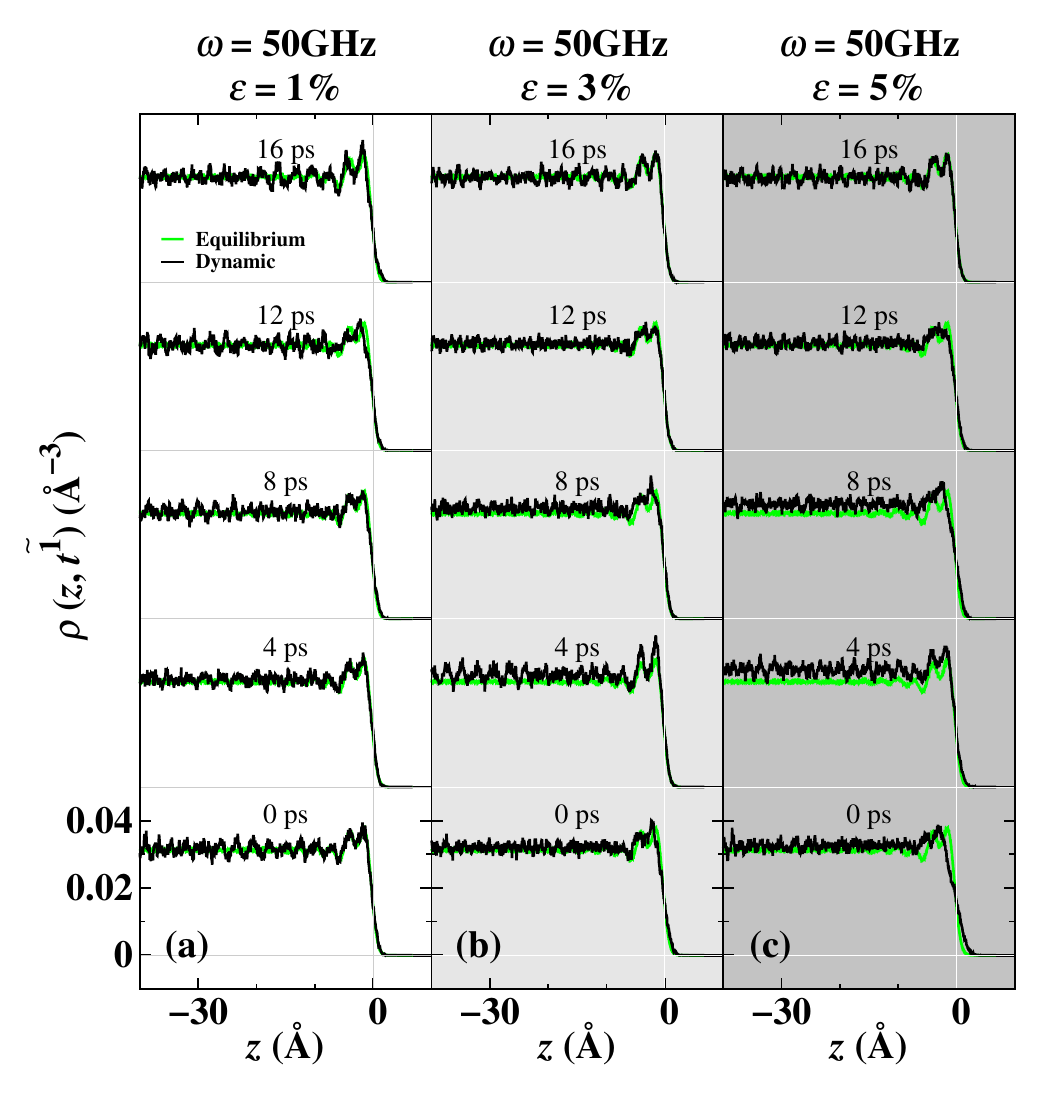}
\caption{The steady oscillation state dynamic fine-grained density profiles at different delay times over one load cycle ($\tilde{t^1}\equiv t \mod C$). Under the cyclic loading frequency of 50GHz, and loading amplitude of $\varepsilon=1\%$ (a), $\varepsilon=3\%$ (b), $\varepsilon=5\%$ (c). Bottommost and topmost curve represents the result for the $\tilde{t^1}=0 ps$ and $\tilde{t^1}=16 ps$, respectively. The curve in green represents the fine-grained density profile of the equilibrium molten Pb surface at $T=T_\mathrm{m}$, which serves as a reference in comparison with the dynamic interfacial profiles.}\label{fig9}
\end{figure}

By contrast, the fact in the corresponding timescale ($C/4=$5 ps) is comparable to the bulk liquid density relaxation time for the highest loading frequency case ($\omega=$50 GHz), suggests a higher probability that the natural atomic-packing relaxation path is altered anisotropically by the applied load, leading to significant adjustments in dynamic density and stress profiles and the birth of non-hydrostatic conditions in bulk liquids. Moreover, because the load is too fast and short in time, the dynamic re-adjustment in the density fields may not have enough time to be sufficient. Thus the system dynamically compromised to the atomic-packing scenarios, which have relatively low free energy yet not as low as the thermodynamic equilibrium state. The above arguments might interpret, to some extent, the reason for the levitation of the mean dynamic surface tension after entering the steady oscillation state. 

In Fig.\ref{fig9}, more variational details in local atomic-packing can be found from the structural adjustment in the dynamic $\rho(z,\tilde{t^1})$ profiles. We observe that, for the higher loading amplitude cases, e.g., $\varepsilon=$3\% and 5\%, there is an increasingly higher probability that the second density peak (next to the outermost density peak or atomic layer) is suppressed or even disappears, e.g., it changes to a weakened shoulder at $\tilde{t^1}=0$ and eight ps under 50GHz 3\% and 5\% loads. At the same time, the densities in the bulk region behind the surface layers could be uniformly levitated or decreased to values that deviate from the equilibrium melt phase density. These observations, in which the adjustments in the dynamic interfacial profiles for the sub-surface region behave differently from the outermost surface layer, also agree with the previous observations in the dynamic stress profiles.

To obtain further quantitative evidence, we separately calculate the local contributions by decomposing the computation of the dynamic surface tension in Eq.(\ref{eq8b}) into two parts\cite{Li22}, $\gamma(\tilde{t^1}) = \gamma_\mathrm{t}(\tilde{t^1})+\gamma_\mathrm{s}(\tilde{t^1}) = \int_{z_\mathrm{lo}}^{z_1(\tilde{t^1})} S(z,\tilde{t^1}) \mathrm{d}z + \left[\int_{z_1(\tilde{t^1})}^{z_\mathrm{hi}} S(z,\tilde{t^1}) \mathrm{d}z - S_l(\tilde{t^1})L_l(\tilde{t^1}) \right] $. $\gamma_\mathrm{t}(\tilde{t^1})$ and $\gamma_\mathrm{s}(\tilde{t^1})$ stand for the contribution of the outermost positive peak and the rest region of the dynamic surface stress profile, respectively. $z_1(\tilde{t^1})$ is the position where the positive stress peak ends at time $\tilde{t^1}$.

\begin{figure}[!htb]
\centering
\includegraphics [width=0.5\textwidth] {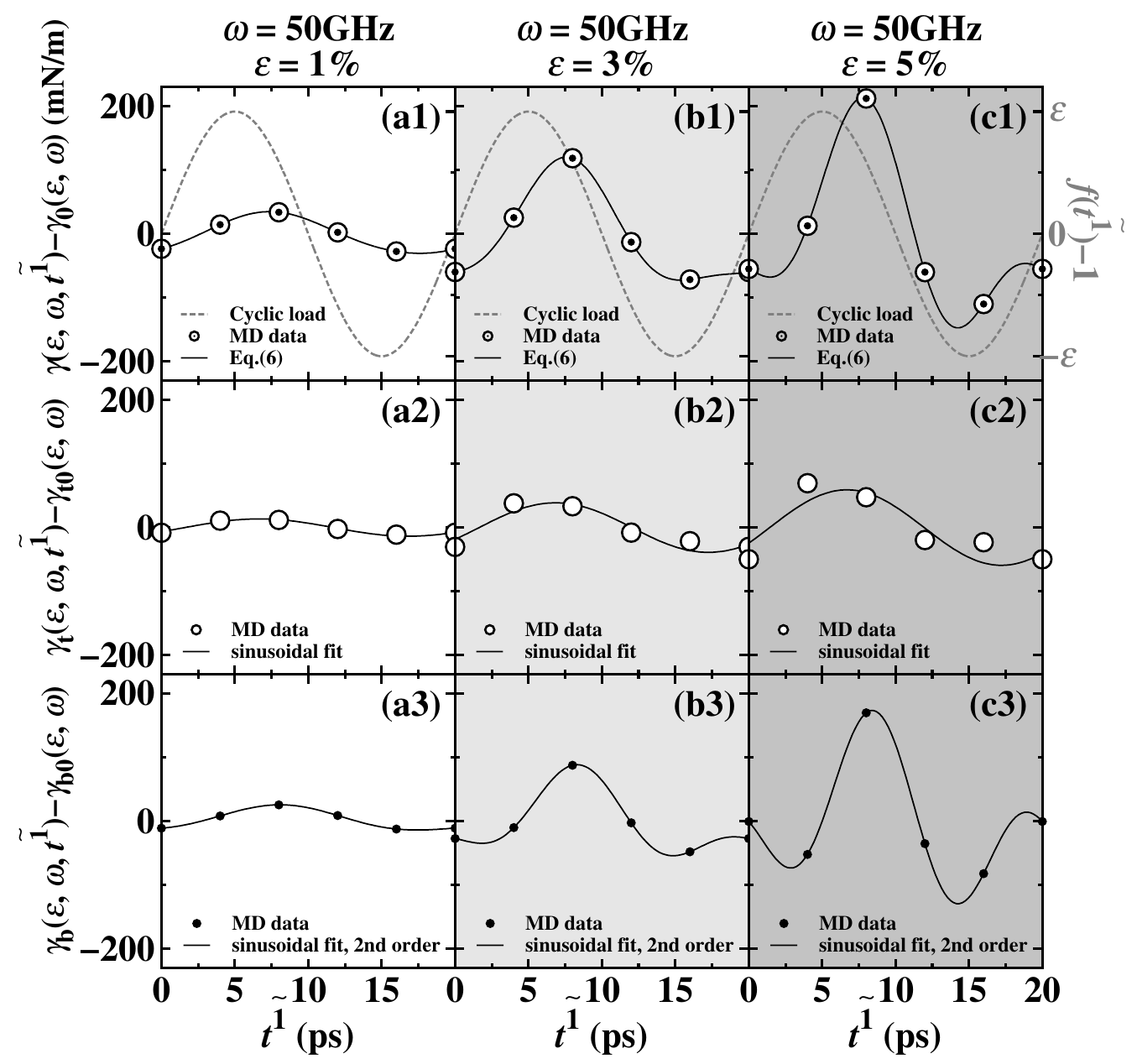}
\caption{Two contributing components of the dynamic surface tension in revealing the steady oscillation state responses of the dynamic surface tension of the molten Pb surfaces to sinusoidal cyclic loads, for the simulation cases under the cyclic loading frequency of 50GHz, and loading amplitude of $\varepsilon=1\%$ (a1-a3), $\varepsilon=3\%$ (b1-b3), $\varepsilon=5\%$ (c1-c3). The open and filled circles represent the contributions of the outermost positive peak and the rest region of the dynamic surface stress profile.}
\label{fig10}
\end{figure}

In Fig.\ref{fig10}, we review the dissected dynamic surface tension variations for the simulation cases under the cyclic loading frequency of 50GHz. It is observed that oscillation of the $\gamma_\mathrm{t}(\tilde{t^1})$ due to the outermost positive peak contribution, well follows the pure sinusoidal function, which has the same frequency as the driven frequency $\omega$, even for the $\varepsilon=5\%$ case. By contrast, the oscillation of the $\gamma_\mathrm{b}(\tilde{t^1})$ results are seen could not be simply described by a pure sinusoidal function with frequency $\omega$. Instead, we need to include the higher frequency sinusoidal function components ($2\omega$), e.g., see in the Fig.\ref{fig10}(a3,b3,c3). These microscopic dissection data support our observations from the above dynamic interfacial density and stress profiles, as well as the speculation that the near-surface layers and sub-layers may respond differently to the applied loads with two different natural frequencies ($\omega_{01}$ and $\omega_{02}$) of the dynamic surface tension oscillation.

Interestingly, one recent study on tuning the dynamic surface tension of molten metals through ultrashort laser pulse irradiation\cite{Li22}, including some authors, has reported similar biased local stress field modification between the outermost surface layer and the liquid interior layers. Within such ultrafast laser modulation, liquids behind the exterior atomic layer receive the deposited laser energy on a time scale comparable to the density relaxation time, while the raw mechanical scenario within the outermost layer remains nearly unaltered. In short, based on the findings from this work and the Ref.\onlinecite{Li22}, one affirms that one consideration must be paid in the modulation of the dynamic liquid surface tension under the ultrafast non-equilibrium condition, i.e., the liquid surface layer and the sub-surface liquid layers are mechanically different and contribute differently to the dynamic surface tension's variation. Further efforts in clarifying the uniqueness of the mechanical and thermodynamical properties of the liquid surface layer\cite{Smith22,Rahman22}, including the nature of the natural frequency and the damping constant mentioned here, are warranted.

\section{Conclusion}
In summary, we design and apply a methodology for computationally investigating the mechanical response of the molten metal surface system to the lateral mechanical cyclic loads via atomistic simulation. By characterizing the dynamic liquid-vapor interfacial stress profile, we predict a potential systematic modulation of the dynamic surface tension of a pure molten metal surface in response to sinusoidal cyclic loads of different frequencies and amplitudes parallel to the surfaces, at a constant temperature.

After entering the steady oscillation state, the oscillation of the dynamic surface tension in response to the applied cyclic load, including the excitation of higher frequency vibration mode at loads with higher driving frequencies and amplitudes, is found to well follows the textbook theory of the driven damped oscillator in classical mechanics. In the meantime, a notable distinction of the liquid surface system from the classical mechanical single-body oscillator system is that the mean values of the steady oscillate state dynamics surface tensions could be levitated significantly, departing from the values of their equilibrium states. For the pure molten metal surface studied, under the highest frequency and amplitude of the applied cyclic load, such levitation could reach $\sim$ 5\% of the equilibrium surface tension. The peak and trough values of the instantaneous dynamic surface tension could reach up to 40\% more and up to 20\% less than the equilibrium surface tension, respectively. 

Two generalized natural frequencies and two generalized damping constants are extracted from the calculated oscillatory data of the dynamic surface tension. Based on the relationship between the natural frequency and the corresponding damping constant, the current system is identified to be underdamped and predicted to experience resonances happen right over the highest driven frequency we applied, i.e., 50GHz. By analyzing the dynamic fine-grained interfacial density and stress profiles, we learn that i) the particle packing density and the local stress adjustments are significantly different from the temperature-induced 
adjustments for the equilibrium liquid-vapor interfaces, in which the whole surface region gets broader and more diffuse as the temperature increases\cite{Li22}, ii) the particle packing adjustments and the local stresses for the outermost surface layers and the sub-surface layers respond differently to the cyclic load, thus contributing differently to the oscillation of the dynamic surface tension. Therefore, we speculate that the outermost surface layers and the sub-surface layers possess distinct natural frequencies, which correlate with the microscopic timescales of the density relaxations at the corresponding regions. Unfortunately, evidence was not sufficient to support a plausible speculation about the nature of the damping constants.

Surface tension is widely recognized to govern various processes and phenomena in fluid dynamics and capillarity, such as droplet and bubble nucleation\cite{Thompson84}, wetting and spreading\cite{deGennes85,Bonn09}, premelting at the metallic surfaces\cite{Tartaglino05}. Knowledge of the variation of the dynamic surface tension of a dynamic surface driven by the applied load obtained here could facilitate the potential tuning of the processes and phenomena controlled by the surface tension. Towards utterly quantitative manipulation of the magnitude of the liquid surface tension and designing material system with proper surface tension variations, more insights and quantitative theories for the natural frequencies and damping constants, as well as the packing structure under a steady oscillation state, are warranted. As natural extensions of the current work, the binary alloy melt surface system should be investigated to examine whether the mechanical theory of the driven oscillator continues to hold and what extra complexity could be caused by the dynamic surface segregations.

\begin{acknowledgments}

YY acknowledges the Chinese National Science Foundation (Grant No. 11874147), the Natural Science Foundation of Chongqing, China (Grant No. cstc2021jcyj-msxmX1144), Open Project of State Key Laboratory of Advanced Special Steel, Shanghai Key Laboratory of Advanced Ferrometallurgy, Shanghai University (SKLASS 2021-10), the Science and Technology Commission of Shanghai Municipality (No. 19DZ2270200, 20511107700) and the State Key Laboratory of Solidification Processing in NWPU (Grant No. SKLSP202105). W.X. acknowledges the financial support of National Science Foundation of China (Grant No.  52003150) and The Program for Young Eastern Scholar at Shanghai Institutions of Higher Learning (Grant No. QD2019006).

\end{acknowledgments}

\bibliography{ref}

\end{document}